\documentclass[aps,prx,twocolumn,superscriptaddress,showkeys]{revtex4-2}
\usepackage{graphicx}
\usepackage{booktabs}
\usepackage[table]{xcolor}
\usepackage{multirow}
\usepackage{amsmath}
\begin{document}

\title{Superconducting $T_\mathrm{c}$ up to 20.6~K in bulk YSi$_2$ and YSi$_2$/Si superlattices due to chemical flattening}

\newcommand{\Peking}{State Key Laboratory for Artificial Microstructures and Mesoscopic Physics, School of Physics, Peking University Yangtze Delta Institute of Optoelectronics, Peking University, Beijing 100871, China}
\newcommand{\Beijing}{Department of Physics and Optoelectronic Engineering Faculty of Science, Beijing University of Technology, Beijing 100124, China}
\newcommand{\Technology}{Physics Department, Beijing Technology and Business University, Beijing 100048, China}
\newcommand{\Geosciences}{School of Mathematics and Physics, China University of Geosciences, Wuhan 430074, China}

\author{Ding-qing Li}\affiliation{\Technology}
\author{Chong Tian}\affiliation{\Peking}
\author{Juan Du}\email{dujuan1121@bjut.edu.cn}\affiliation{\Beijing}
\author{Jun-jie Shi} \email{jjshi@pku.edu.cn} \affiliation{\Peking}
\author{Pei-song~He}\affiliation{\Technology}
\author{Deng-hui~Xu}\affiliation{\Technology}
\author{Hong-xia Zhong}\affiliation{\Geosciences}
\author{Yao-hui Zhu}
\email{zhuyaohui@th.btbu.edu.cn} \affiliation{\Technology}

\date{\today}

\begin{abstract}
Currently, the fundamental building blocks of leading quantum computers are Josephson junctions, whose core is usually the superconducting Al on Si wafers. However, the transition temperature $T_\mathrm{c}$ of bulk Al ($\sim1.1$~K) is below the boiling point of liquid helium ($\sim4.2$~K), which is one of the challenges to its widespread application. Here, we propose a Si-matched AlB$_2$-type superconductor YSi$_2$ as a promising alternative to Al. The solution of anisotropic (isotropic) Migdal-Eliashberg equation without (with) anharmonicity gives $T_\mathrm{c}\sim20.6$~K (17.2~K), which is at the highest level in silicides. Its excellent superconductivity can be attributed mainly to the Si honeycombs, which become plane here due to the `chemical flattening' effects of the Y atoms instead of being buckled in most silicides. We tested its thermodynamical, kinetic, dynamical, and mechanical stability by first-principles calculations. Particularly, the negative elastic stiffness constant $C_{66}$ calculated by usual methods turns positive even without the zero-point energy once the Si honeycombs are compressed below a threshold. This strain can also make its calculated lattice constants agree with the experimental ones. We propose structures to realize this strain, i.e., YSi$_2$(0001)/Si(111) superlattices, which can also strengthen the overall stability of YSi$_2$ while maintaining its $T_\mathrm{c}$ above 7.0~K.
\end{abstract}

\keywords{Silicide superconductor, Migdal-Eliashberg equation, Anharmonicity, Zero-point energy}

\maketitle

\section{Introduction}

Quantum computing can leverage quantum mechanics to execute particular computational operations that are beyond the reach of even the most advanced classical computers~\cite{QuantumComputing,QuantumComputational}. Thus, research in quantum computing has become prevalent worldwide. Currently, quantum computing employs several modalities, five of which are the most significant: trapped ions, neutral atoms, silicon spin, photon-based, and superconducting qubits~\cite{trappedion,phonon,PhotonQC,qc}. Extensive studies and experiments have confirmed that superconducting qubits are the most accessible and fastest method for achieving quantum computing up to now.~\cite{SuperconductingQubits,SuperconductingQuantumgood,SQ} These qubits are based on the Josephson junctions, and the most mainstream superconductor used here is the superconducting Al on a silicon wafer.~\cite{AlSuperconductingQ} Moreover, Josephson junctions are fundamental building blocks of many other types of quantum electronic systems, e.g., Josephson junction field effect transistors (JJFET).~\cite{Xiong2026} Evidently, the characteristics of the superconductors used here are of significant importance for the performance of these quantum devices.

Si-matched superconductors occupy a critical position in quantum computers and devices.~\cite{Xiong2026} While serving as one of the most abundant elements found on earth, silicon has also displayed outstanding stability, and facilitates the manufacture of quantum circuits due to the mature technology of Si industry. However, the most widely used superconductor on Si wafers, i.e., Al, has a transition temperature of only $\sim1.1$~K in its bulk phase, which is even lower than the boiling point ($\sim4.2$~K) of liquid helium. This makes its widespread application quite difficult. Therefore, it is imperative to find superconductors that can be well matched with silicon wafers and have transition temperatures as high as possible, especially above 4.2~K.

The family of silicides, especially the binary disilicides MSi$_2$ (M is a metal atom) with the AlB$_2$-type structure, have a great potential to yield excellent superconductors exhibiting relatively high $T_\mathrm{c}$ (e.g., above 4.2~K) and matching well with Si substrates.~\cite{scindopedsp3semi,si-ge-c-basedSuperconductors,scSiliconReview} Although CaSi$_2$ in the trigonal phase has a rather low $T_\mathrm{c}$ of 30~mK and its tetragonal phase has a slightly higher $T_\mathrm{c}$ of 1.56~K at ambient pressure, its $T_\mathrm{c}$ could be increased to 14~K when the AlB$_2$-like phase with almost planar Si honeycomb was obtained by applying high pressure ($>16$~GPa).~\cite{sanfilippo2000,Bordet2000} A joint experimental and theoretical study also reported that $T_\mathrm{c}$ of BaSi$_2$ was enhanced when the buckled Si layers were flattened by a high-pressure synthesis, and this mechanism was explained by their calculations based on the density function theory (DFT).~\cite{Livas2011} However, it is challenging to maintain an external high pressure in practical use, especially quantum circuits. Alternatively, it may be easier to make the Si honeycomb flat by using a chemistry method (called `chemical flattening' by us), for example, using trivalent elements like Y instead of Ca or Ba.

In this work, we will show that, among the various Si-based superconductors, AlB$_2$-type YSi$_2$ is a stable and Si-matched candidate with potential $T_\mathrm{c}$ around 20.6~K (above the liquid hydrogen boiling point $\sim20.4$~K). Although many Si-based superconductors have been studied, the highest $T_\mathrm{c}$ is about 17~K, which is found in A15 V$_3$Si. However, it is still challenging to integrate it well with Si wafers due to the formation of impurity phases, such as VSi$_2$, which is not superconducting.~\cite{Zhang2021} The search of new Si-based superconductors leads us to the AlB$_2$-type YSi$_2$. The synthesis of AlB$_2$-type YSi2 was first reported in the 1960s, and its crystal structure as well as related properties were also studied.~\cite{DimorphismRareEarth} Later, several groups also reported the epitaxial growth of YSi$_{1.7}$ and YSi$_{1.9}$ (defected AlB$_2$-type) on the Si(111) surface,~\cite{EpitaxialGrowthRareearth,Epitaxialysi111} the polycrystalline thin films of YSi$_{2-x}$ on Si(111) substrates,~\cite{ThinFilmssi}, the buried YSi$_{1.7}$ by ion-beam synthesis,~\cite{Alford1989,Wu1998} and the growth of nanowires of YSi$_2$ on suitable Si substrates.~\cite{Zeng2008,Iancu2009,Iancu_2013,Formationnanosi110} These experimental results demonstrated that YSi$_2$ matches well with Si substrates in various ways. Unfortunately, the superconductivity of YSi$_2$ has not been investigated even theoretically so far, and this is one of the motivations for our present work.

We focused on two aspects of YSi$_2$, i.e., evaluating its $T_\mathrm{c}$ and testing its overall compatibility with Si substrates by building YSi$_2$(0001)/Si(111) superlattices. On one hand, we calculated $T_\mathrm{c}$ of bulk YSi$_2$ by three approaches, including the McMillan-Allen-Dynes (McAD) formula and the anisotropic (isotropic) Migdal-Eliashberg (ME) equations without (with) the anharmonicity. Its excellent superconductivity can be attributed mainly to the Si honeycombs, which become plane here due to the `chemical flattening' effects of the Y atoms instead of being buckled in most silicides. On the other hand, although the synthesis of YSi$_2$ has made significant progress, there are still some controversial issues that need be addressed. Therefore, we first tested its thermodynamical, kinetic, dynamical, and mechanical stability by first-principles calculations. Then we proposed that a proper compressive strain in the planes of the Si honeycombs could turn the calculated elastic stiffness constant $C_{66}$ positive and also make the calculated lattice constants of YSi$_2$ agree with the experimental ones. One of the possible ways to realize this strain is the formation of YSi$_2$(0001)/Si(111) interfaces.~\cite{ThinFilmssi} This motivate us to build YSi$_2$(0001)/Si(111) superlattices, which can also improve the overall stability of YSi$_2$ and demonstrate its atomic-level compatibility with Si substrates while maintaining its $T_\mathrm{c}$ above 4.2~K. Moreover, this kind of superconducting multilayers can be applied in many devices, such as single-photon detectors.~\cite{Wu2026}

\section{Results and Discussion}

\begin{figure}
\includegraphics[width=0.48\textwidth]{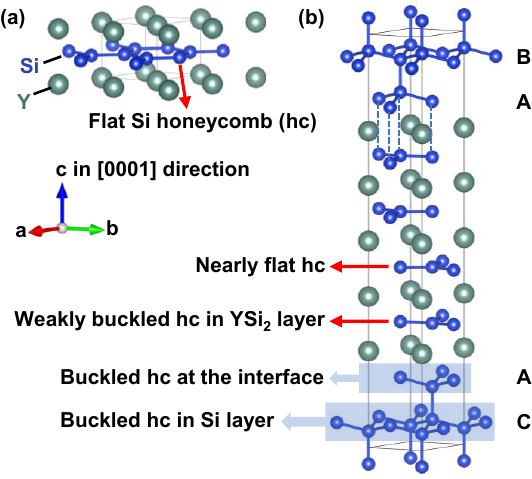}%
\caption{\label{structure}(a) The crystal structure of bulk AlB$_2$-type YSi$_2$. Its unit cell is repeated twice in the \textbf{a} and \textbf{b} directions, i.e., 2$\times$2$\times$1, to illustrate the flat Si honeycomb clearly. (b) The crystal structure of an YSi$_2$(0001)/Si(111) superlattice. For example, in the unit cell of SL1, the middle part includes 5 layers of Y atoms and 4 Si honeycombs (2 weakly buckled and 2 nearly flat) stacked alternately along the \textbf{c} direction. The rest part comprises 4 buckled Si honeycombs in the ABC stacking order (or diamond-like), and 2 of them are at the interfaces. The four vertical dashed lines indicate that the buckled Si honeycombs at the interface have the same in-plane atomic coordinates as the ones inside the YSi$_2$ part.}
\end{figure}

In this section, we will present and discuss the results of first-principles calculations for the AlB$_2$-type YSi$_2$ in three scenarios. The strainless bulk YSi$_2$ is examined comprehensively as a reference. Then a compressive stress is applied to the bulk YSi$_2$ in the Si honeycomb plane while it is free of any stress in the out-of-plane direction. The reasons for this will be given below. Finally, we construct a series of YSi$_2$(0001)/Si(111) superlattices, which can realize this in-plane compressive strain and demonstrate the compatibility of YSi$_2$ with Si as well.

\subsection{Structure and Stability}

\subsubsection{Strainless bulk YSi$_2$}

We start from the bulk structure of the AlB$_2$-type YSi$_2$ free of any strains as shown in Fig.~\ref{structure}(a). It crystallizes in the hexagonal crystal system and has the space group P6/mmm (no.~191). Our first-principles calculations show that the in-plane and out-of-plane lattice constants are $a=4.07$~Å  and $c=3.89$~Å, respectively, which are consistent with the calculations for the same structre done by other groups, e.g., Ref.~\onlinecite{Magaud1997} and Materials Project.~\cite{osti_1200881} Even at the ambient pressure, the Si honeycombs are perfectly plane due to the stronger `chemical flattening' effects of the Y atoms than the Ca or Ba atoms,~\cite{sanfilippo2000,Bordet2000,Livas2011} and this will be illustrated from another viewpoint by using the YSi$_2$/Si superlattices below. Moreover, the electron localization function indicates that the electrons are rather delocalized and hence the metallic bond is dominant in YSi$_2$ as shown in Fig.~1 of the Supplemental Information (SI).

\begin{figure}
\includegraphics[width=0.49\textwidth]{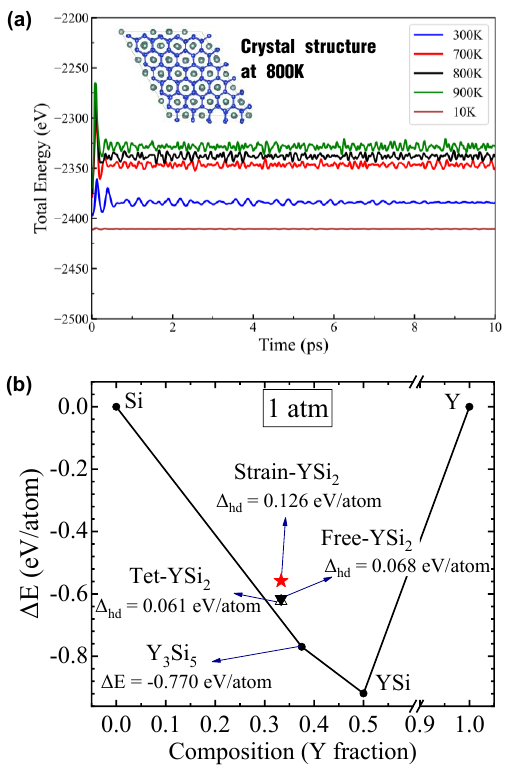}%
\caption{\label{stability}(a) The NVT AIMD simulation. The total energy is plotted as a function of time at the temperatures of 10~K (brown line), 300~K (blue line), 700~K (red line), 800~K (black line), and 900~K (green line). The inset shows the crystal structure along c axis after NVT AIMD simulation at 800~K. (b) The convex hull of Yi-Si system. The red star stands for the strained YSi$_2$ (labeled by `Strain-YSi$_2$'), which has the same structure as the strainless YSi$_2$ (`Free-YSi$_2$', filled triangle-down marker) but different lattice constants. The open triangle-up marker denotes the tetragonal YSi$_2$ (`Tet-YSi$_2$'), whose structure is taken from the Open Quantum Materials Database (OQMD, ID: 19375). The structure of Y$_3$Si$_5$ has the space group P$\bar{6}$2m (no.~189), which is the same as that in Ref.~\cite{Magaud1997} and can also be found in OQMD (ID:~1795316). The structure of YSi (space group Cmcm or no.~63) is taken from the Materials Project (ID: mp-9972).}
\end{figure}

Using the \textit{ab initio} molecular dynamics (AIMD) simulation, we tested the kinetic stability of bulk YSi$_2$. The results based on the NVT ensenmble at 10~K, 300~K, 700~K, 800~K and 900~K are shown in Figure~\ref{stability}(a). It can be seen that the total energy fluctuation falls in a small range usually after the simulation lasts 1~ps. The inset in Figure~\ref{stability}(a) shows the crystal structure after the NVT simulation of 10~ps at 800~K. It is obvious that the structure still maintain relative stability although atoms have undergone minor displacements. Extensive NVT AIMD simulations show that the structure is stable up to 1800~K and the associated details can be found in the SI. These results confirm that YSi$_2$ crystals have good kinetic stability.

We assessed the thermodynamical stability of bulk YSi$_2$ using the convex hull and including the zero-point energy (ZPE) correction as shown in Fig.~\ref{stability}(b). The computation method is detailed in the SI. We particularly calculated the formation energy of bulk YSi$_2$ in the tetragonal phase (space group I4$_1$/amd or no.~141), in the strainless AlB$_2$-type phase, and in the strained AlB$_2$-type phase (the in-plane lattice constant compressed to 3.91~Å). The formation energy of strainless AlB$_2$-type YSi$_2$ ($\Delta{E}=-$0.616~eV/atom) is slightly higher than that of the tetragonal structure ($\Delta{E}=-$0.623~eV/atom). The AlB$_2$-type YSi$_2$ (0.033~eV/atom) has smaller ZPE correction than the tetrogonal one (0.036~eV/atom), which make their formation energy closer. Meanwhile, both of them are above the line connecting elemental Si and Y$_3$Si$_5$. Their vertical distances to the line are called hull distances $\Delta_\mathrm{hd}$ or energy above hull, which describe how poor the stability is. Their hull distances are shortened by the ZPE correction of Y$_3$Si$_5$ (0.039~eV/atom), which is the largest in these compounds. Both of the distances are smaller than 0.1~eV/atom as shown in Fig.~\ref{stability}(b), and thus they can be regarded as good metastable structures.~\cite{Ma2017}

\begin{figure*}
\includegraphics[width=1.0\textwidth]{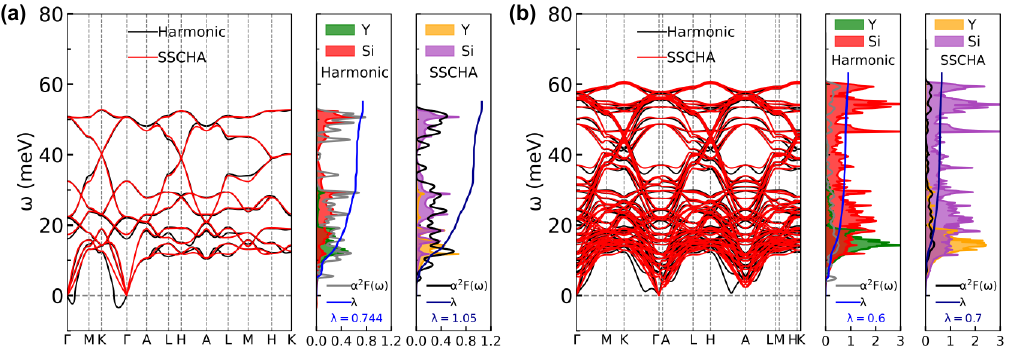}%
\caption{\label{elph}(a) The phonon spectra (left panel) of strainless bulk YSi$_2$. The black (red) curves depict the harmonic (anharmonic) dispersions. The anharmonicity is taken into account by the SSCHA package. The projected phonon DOS (PHDOS), Eliashberg spectral functions $\alpha^2F(\omega)$, and integral EPC constant $\lambda$ are presented separately for harmonic (central panel) and anharmonic (right panel) cases. (b) The same as (a) but for the YSi$_2$/Si superlattice SL1.}
\end{figure*}

We also tested the dynamical stability of bulk YSi$_2$ by calculating its phonon spectra with and without the anharmonicity. Figure~\ref{elph}(a) shows that its lowest phonon branch has small negative frequencies near point $\Gamma$. This branch looks similar to the out-of-plane acoustic (ZA) phonon mode of two-dimensional materials, whose frequency is proportional to the square of wavevector and usually has negative frequency due to numerical errors. This is reasonable because the layered structure of YSi$_2$, especially its Si honeycomb, also bears similarity to two-dimensional matersials. Furthermore, the negative frequencies in this phonon branch disappear when the anharmonicity is taken into account by the Stochastic Self-Consistent Harmonic Approximation (SSCHA) as shown in Fig.~\ref{elph}(a). Thus the calculated phonon spectra support the dynamical stability of bulk YSi$_2$.

The mechanical stability of bulk YSi$_2$ was tested by calculating the elastic stiffness tensor. Calculations on the DFT level with dense k-mesh (denser than 20$\times$20$\times$20) show that, when the bulk YSi$_2$ is free of any stress, its elastic stiffness constant $C_{66}$ is slightly negative indicating that the structure is not stable enough. But the usual calculations do not include the ZPE correction, which is remarkable in YSi$_2$. Thus we calculated $C_{66}$ again using the energy-strain method and taking into account the ZPE correction. This gives us $C_{66}=2.2$~GPa, which shows that the bulk YSi$_2$ is also stable mechanically although `soft' under in-plane shear. The complete elastic stiffness tensor is presented in Table~1 of the SI. Our detailed calculations also show that $C_{66}$ could become positive even without the ZPE correction, when $a$ was reduced below a critical value $\sim3.98$~Å and $c$ was allowed to increase freely (see details below). We will argue below that this kind of strain might have been present in the samples prepared by some experiments and made the structure more stable mechanically.

\subsubsection{Strained bulk YSi$_2$}

The effects of strain on $C_{66}$ motivates us to test the stability of the bulk YSi$_2$ subjected to an in-plane compressive stress in more details. To be specific, we took the strained YSi$_2$ with an in-plane lattice constant of 3.91~Å as an example. In parallel to the strainless bulk YSi$_2$, we also did AIMD simulation, calculated the phonon spectra, the formation energy and the elastic stiffness tensor. The NVT AIMD simulation shows that the structure is stable up to 1000~K. The lowest phonon branch also has small negative frequencies, which can be removed by including the anharmonicity. The formation energy of the compressed YSi$_2$ is $-0.558$~eV/atom, whose hull distance is 0.126~eV/atom as shown in Fig.~\ref{stability}(b). As for the elastic stiffness tensor ($C_{ij}$) of YSi$_2$, only five terms are required to completely describe the mechanical behavior of hexagonal materials as given in Table~2 of the SI. Mechanical stability requirements for hexagonal systems apply the following restrictions on their elastic stiffness constants:
$C_{11}>|C_{12}|$, $2 C_{13}^{2}<C_{33}\left(C_{11}+C_{12}\right)$, $C_{44}>0$, and $C_{66}>0$.~\cite{Mouhat2014} The data we have obtained satisfy these constraints, indicating that the strained YSi$_2$ exhibits mechanical stability. The calculated Young's modulus (E) is 138~GPa. The ductility or brittleness of a material is typically determined by using the Pugh's modulus ratio (B/G) and the Poisson's ratio (v). A higher Pugh's ratio indicates greater ductility and less brittleness in the material (see more details in the SI). The critical value that distinguishes ductility from brittleness is 1.75 for the B/G ratio. Moreover, the Poisson's ratio (v) should be larger than or equal to 0.26 for ductile materials, otherwise the material exhibits brittleness. The calculations based on Table~2 of the SI show that the strained YSi$_2$ has a B/G ratio of 3.22 and its Poisson's ratio (v) is 0.36. According to the criteria above, bulk YSi$_2$ exhibits excellent ductility. In summary, we can still regard the compressed YSi$_2$ as a metastable structure.

\begin{figure}
\includegraphics[width=0.5\textwidth]{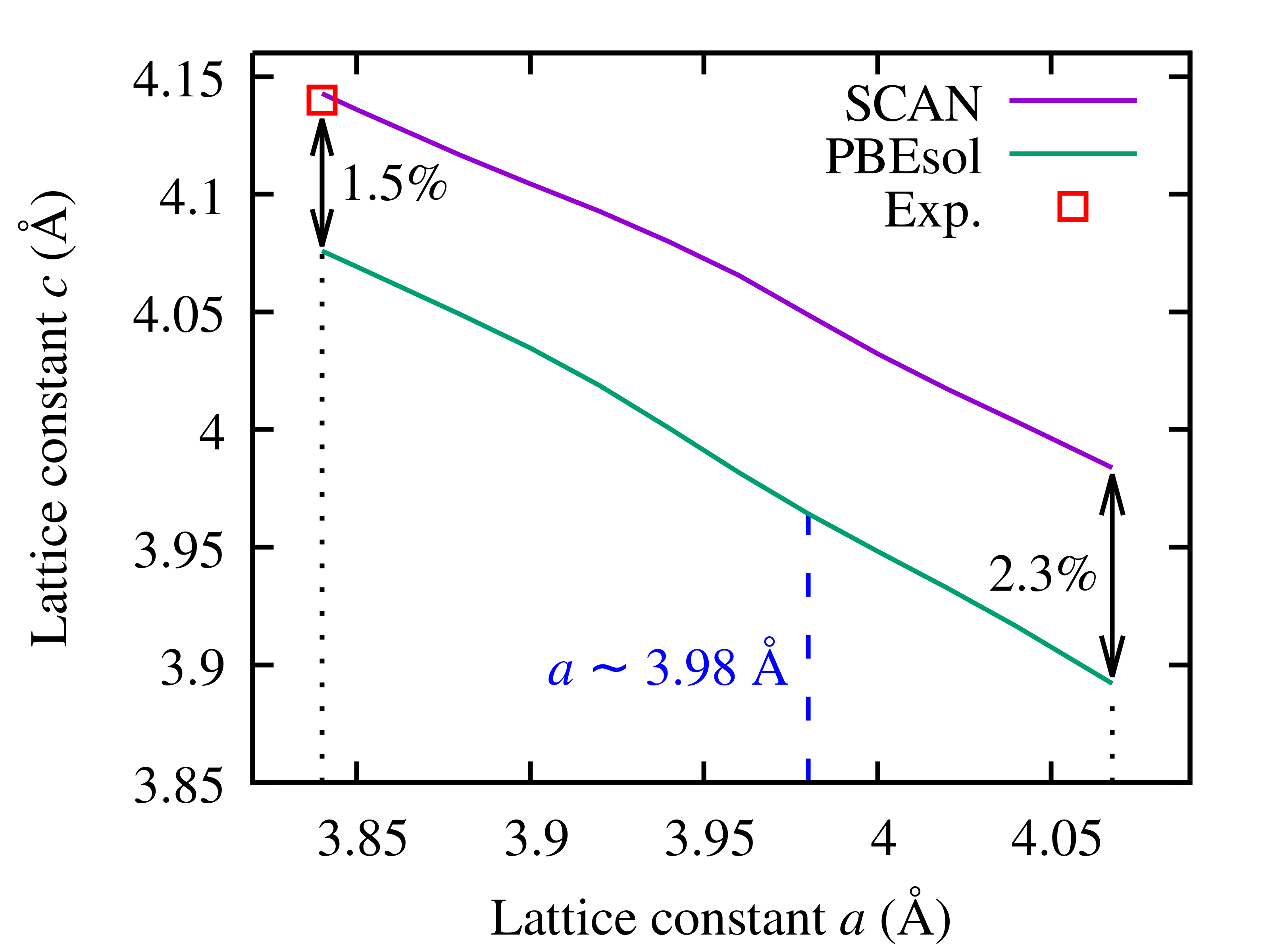}%
\caption{\label{strain} The variation of $c$ with $a$. In the strainless state, $a=4.07$~Å and $c=3.89$~Å are determined by the PBEsol functional. For the same $a=4.07$~Å, the SCAN functional yields $c=3.98$~Å, which is larger than the PBEsol one by 2.3\%. When $a$ was compressed below a critical value $\sim3.98$~Å, $C_{66}$ would turn positive even without the ZPE correction. The red square denotes the experimental lattice constants, i.e., $a=3.842$~Å and $c=4.14$~Å. The experimental $c$ is larger than the corresponding PBEsol result by only 1.5\%, and almost identical to that given by the SCAN functional.}
\end{figure}

Next we will compare our calculated lattice constants with the experimental ones. The calculated $a=4.07$~Å is larger than its corresponding experimental value $\sim3.842$~Å while the calculated $c=3.89$~Å is smaller than the experimental one $\sim4.14$~Å as reported by several groups.~\cite{structurevariation,EpitaxialGrowthRareearth,RAREEARTHDISILICIDES} The difference between the theoretical and experimental values is remarkable, and this issue was investigated by several groups.~\cite{Magaud1997,Rogero2002,AlB2vac} In order to explain the experimental results of defected YSi$_2$, i.e., YSi$_{1.7}$, epitaxially grown on Si(111) substrate by rapid heating with an electron beam (producing a peak temperature of 1200-1400 K),~\cite{EpitaxialGrowthRareearth} these groups introduced one Si vancancy in the six-membered ring of Si atoms. Then they constructed Th$_3$Pd$_5$-type or Yb$_3$Si$_5$-type structures of Y$_3$Si$_5$ from the $\left(\sqrt{3}\times\sqrt{3}\right)R30^{\circ}$ supercell of the AlB$_2$-type YSi$_2$. As these model structures of Y$_3$Si$_5$ have different symmetries and unit cells from those of the AlB$_2$-type YSi$_2$, they managed to convert the lattice constants of Y$_3$Si$_5$ back to those of YSi$_2$. In this way, they achieved the agreement between the calculated and experimental lattice constants. However, these groups did not discuss other experimental results of the AlB$_2$-type YSi$_2$ with an Y:Si mole ratio of 1:2. These YSi$_2$ powder were  prepared by different techniques at relatively low temperatures (e.g., 1100~$^\circ$C in Ref.~\onlinecite{RAREEARTHDISILICIDES} and 500~$^\circ$C in Ref.~\onlinecite{DimorphismRareEarth}). Obviously, it is also necessary to explain these experimental results.

To this end, we introduce a new interpretation of the discrepancy between the experimental and the calculated lattice constants: it may arise from an in-plane compressive strain present in the synthesized samples. The possible sources of the strain will be discussed later. This interpretation are supported by two sets of calculations done by us. We first studied the variation of the structure with an in-plane compressive strain using two kinds of functionals. The PBEsol functional in Fig.~\ref{strain} shows that, when $a$ is reduced to the experimental value 3.842~Å by an in-plane compressive stress alone, $c$ increases to 4.08~Å that is smaller than the experimental value, i.e., 4.14~Å, by only 1.5\%. However, $c$ calculated by the Strongly Constrained and Appropriately Normed (SCAN) semilocal density functional is in perfect agreement with the experimental result under the same strain. The results given by the SCAN functional are usually closer to experimental data. Secondly, a small in-plane compressive strain strengthens effectively the mechanical stability as pointed out above. Thus a proper in-plane compressive strain can resolve both of the problems at the same time.

\subsubsection{YSi$_2$(0001)/Si(111) superlattices}

Among various sources of the in-plane compressive strain, one possibility is the interface formed between YSi$_2$(0001) and Si(111) lattice planes as shown in Fig.~\ref{structure}(b).~\cite{ThinFilmssi} The Si(111) lattice planes can be treated as buckled honeycombs of Si, and its lattice constant is around 3.84~Å, which is close to both the experimental ($\sim3.842$~Å) and theoretical ($\sim4.07$~Å) results of $a$ in YSi$_2$. The buckled honeycomb of Si(111) surface has an obvious correspondence to the flat honeycomb of Si in YSi$_2$ as illustrated by the dashed vertical lines in Fig.~\ref{structure}(b). Moreover, Si phase could not be ruled out in these experiments,~\cite{RAREEARTHDISILICIDES} and thus YSi$_2$ can be grown on Si(111) surface layer by layer along the [0001] direction. Due to the formation of the interface, the in-plane lattice constant $a$ ($\sim4.07$~Å) of YSi$_2$ will be compressed at least in the region close to the interface. At the same time, the buckled honeycombs of Si(111) lattice planes ($a\sim3.84$~Å) are expanded and deviate from those in the bulk Si crystal. 

To demonstrate quantitatively the role played by the YSi$_2$/Si interface, we constructed a series of YSi$_2$(0001)/Si(111) superlattices, where only such kind of interface is present. Figure~\ref{structure}(b) shows the structure of the simplest superlattice (labeled by SL1) in our study, composed of nearly 5 periods of YSi$_2$ (including 4 nearly flat Si honeycombs) in the (0001) direction and 4 buckled Si(111) layers. In order to make the symmetry of the superlattices as high as possible, we increase the number of Si(111) planes by three since its buckled honeybomb repeats itself every three layers (due to the ABC stacking of the fcc lattice) starting from the minimum of 4 layers. The YSi$_2$ periods in the (0001) direction can be increased one by one. The highest symmetry that we can get has the space group number of 164, or $P\bar{3}m1$, which belongs to the trigonal crystal system and the hexagonal lattice system at the same time. To see the effects of layer thickness, we also calculated another two superlattices: one (labeled by SL2) composed of nearly 5 periods of YSi$_2$ and 7 buckled Si(111) layers, and the other (labeled by SL3) composed of nearly 8 periods of YSi$_2$ (including 7 nearly flat Si honeycombs) and 4 buckled Si(111) layers. More details are given in Fig.~2 of the SI.

\begin{figure*}
\includegraphics[width=1.0\textwidth]{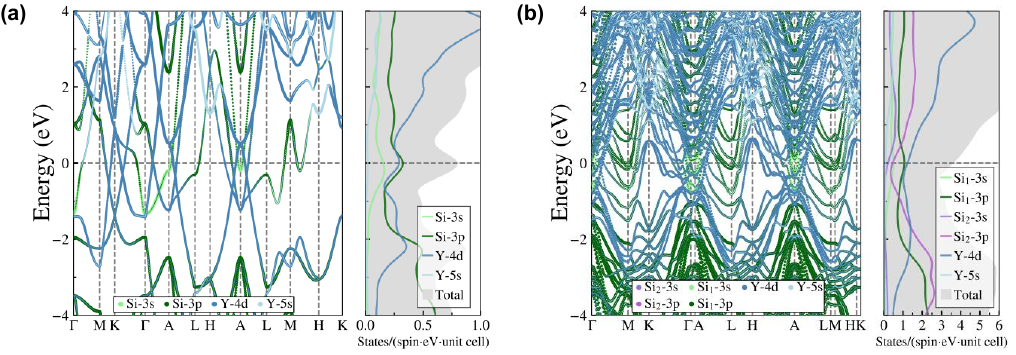}%
\caption{\label{band}(a) The projected band structure and DOS of strainless bulk YSi$_2$. In the band structure, the sizes of the colored circles are proportional to the contribution of their corresponding orbitals. (b) The projected band structure and DOS of the YSi$_2$/Si superlattice SL1. In the DOS diagram, Si$_1$ and Si$_2$ stand for the Si atoms in the YSi$_2$ part and diamond-like Si part of their superlattice, respectively.}
\end{figure*}

Here we take SL1 in Fig.~\ref{structure}(b) as an example to show the fundamental characteristics. Before the relaxation, the lattice constant of Si(111) plane is 3.84 Å, aligning with the experimental value. Following the structural optimization, the lattice constant rose from 3.84~Å to 3.97~Å. This change could be attributed to the larger lattice constant $a$ of YSi$_2$ in comparison to the unrelaxed Si(111) planes. When these materials are combined, the YSi$_2$ layers extend the Si(111) planes, while the Si(111) planes compress the in-plane lattice constant of YSi$_2$, ultimately resulting in an equilibrium lattice constant of 3.97~Å. Our calculations also show that $a$ decreases to 3.94~Å in SL2, which has thicker Si(111) layers. It is natural that the equilibrium lattice constant $a$ decreases with the number of buckled Si(111) layers, and approaches the experimental value of 3.842~Å when the Si layer is much thicker than the YSi$_2$ layer in the superlattices. Comparing the Si honeycombs in Fig.~\ref{structure}(b), one can easily find that the ones in the YSi$_2$ portion are flatter than those in the Si portion, which illustrates the `chemical flattening' of Y atoms from another viewpoint.

To test the stability of YSi$_2$/Si superlattices, we conducted NVT AIMD simulations and calculated their phonon spectra as well as elastic stiffness tensors. The AIMD simulation results, as shown in the figures of the SI, indicate that the structures are still stable at 1000~K and do not change their original configurations. Moreover, the calculated phonon spectra of SL1 is shown in Fig.~\ref{elph}(b), and the corresponding results for SL2 and SL3 are presented in the SI. It is worthwhile to point out that the phonon spectra of the superlattices have no negative frequencies even if the anharmonicity is not taken into account. This suggests that the superlattices are rather stable dynamically. Finally, Tables~3 of the SI lists the calculated elastic stiffness tensors of the three superlattices, which demonstrate their mechanical stability.

\subsection{Electronic structure, electron-phonon coupling and superconductivity}

The projected band structure and density of states (PDOS) are shown in Fig.~\ref{band}(a) for the strainless bulk YSi$_2$. The Fermi level coincides with a local maximum of DOS, which is usually beneficial to superconductivity. The Si 3p and Y 4d orbitals almost have equal contributions to the DOS at the Fermi level, and the contribution of the Si 3s orbitals to DOS is nearly half of that of the Si 3p orbitals. The sum of the Si 3s and 3p orbitals dominates the DOS at the Fermi level, and this indicates that Si may play a crucial role in the superconductivity. Although the contribution of the Y 5s orbitals to DOS is relatively small, it has a noticeable value.  Consequently, these may make the strainless bulk YSi2 a four-band (at least three-band) superconductor, which will be shown in detail later. When an in-plane compressive stress is applied to YSi$_2$, the contribution of the Y 5s orbitals becomes even smaller as shown in Fig.~3 of the SI and it may become a typical three-band superconductor.

\begin{table*}
\caption{\label{Tc-table}The superconducting transition temperature $T_\mathrm{c}$ (in the unit of K) of the strainless bulk YSi$_2$ calculated by using four methods and two different sets of q-mesh (k-mesh). The results of anharmonic McAD and isotropic ME were both given by the SSCHA package. Moreover, $\mu^\ast=0.10$ is used in the solution of both isotropic and anisotropic ME equations.}
\begin{ruledtabular}
\begin{tabular}{ccccccc}
    \multirow{2}{3em}{\textbf{q-grid}} & \textbf{Coarse} & \textbf{Fine} & \textbf{Harmonic} & \textbf{Anharmonic} & \textbf{Anharmonic} & \textbf{Harmonic} \\
     & \textbf{k-mesh} & \textbf{k-mesh} & \textbf{McAD} & \textbf{McAD} & \textbf{Isotropic ME} & \textbf{Anisotropic ME} \\ \hline
    5$\times$5$\times$5 & 10$\times$10$\times$10 & 20$\times$20$\times$20 & 8.9 & 15.7 & 16.8 & 20.6\footnotemark[1] \\ 
    6$\times$6$\times$6 & 12$\times$12$\times$12 & 24$\times$24$\times$24 & 8.1 & 14.2 & 15.3 & 20.6\footnotemark[2] \\
\end{tabular}
\end{ruledtabular}
\footnotetext[1]{The solution of the anisotropic ME equation used a fine q-mesh ($25\times25\times25$) and k-mesh ($50\times50\times50$).}
\footnotetext[2]{The solution used a fine q-mesh ($30\times30\times30$) and k-mesh ($60\times60\times60$).}
\end{table*}

\begin{figure*}
\includegraphics[width=1.0\textwidth]{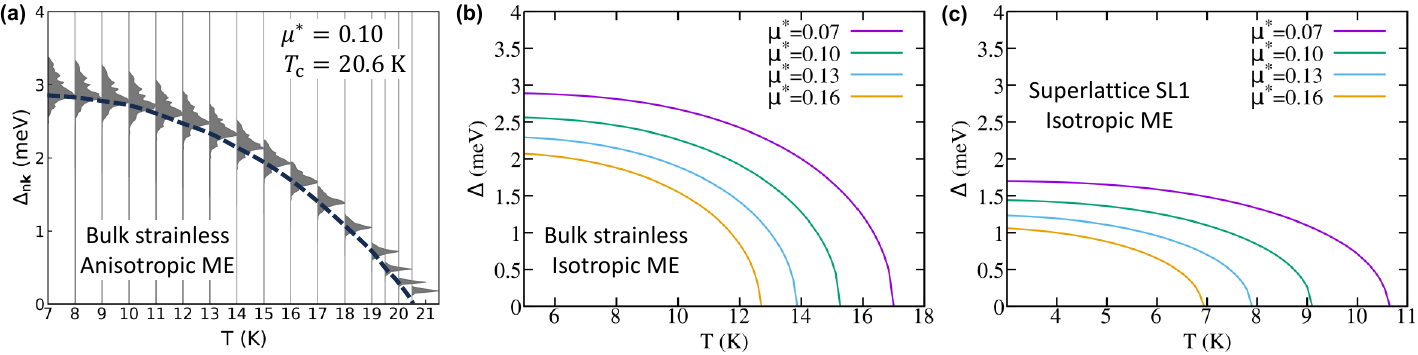}%
\caption{\label{Tc}(a) The anisotropic superconducting gap $\Delta_{n\bf{k}}(\omega=0)$ of the strainless bulk YSi$_2$ as a function of temperature. At each temperature, the filled curve indicates the number of states on the Fermi surface with the superconducting gap. The dashed line is a fitting curve for the peak positions of $\Delta_{n\bf{k}}(\omega=0)$, and it drops to zero around $T_\mathrm{c}=20.6$~K. This calculation used a q-mesh of 6$\times$6$\times$6 and corresponding k-mesh as given in Table~\ref{Tc-table} and its footnote~b. The effective Coulomb potential is set to $\mu^\ast=0.10$ as usual. (b) The isotropic superconducting gap of the strainless bulk YSi$_2$ as a function of temperature for several typical $\mu^\ast$. The q- and k-mesh used in the calculations are listed in Table~\ref{Tc-table}. (c) The same as (b) but for the superlattice SL1. The q- and k-mesh used in the calculations are listed in Table~\ref{Tc-table-sl}.}
\end{figure*}

The YSi$_2$/Si superlattices have much more complicated electronic structures, and the one for SL1 is presented in Fig.~\ref{band}(b). In general, the superlattices are still metallic at least in the in-plane directions. Although the Fermi level does not coincide with the local maximum of DOS, it is still quite large and this may also give rise to remarkable superconductivity. To distinguish the contributions of the Si atoms from the YSi$_2$ and Si parts, we projected the total DOS of Si atoms separately. The Si orbitals from the YSi$_2$ part, denoted by Si$_1$-3s and Si$_1$-3p, dominate the DOS at the Fermi level. On the contrary, the Si orbitals from the diamond-like Si part, denoted by Si$_2$-3s and Si$_2$-3p, have smaller DOS (near the local minimum) at the Fermi level. This suggests that the Si honeycombs of the YSi$_2$ part may still play the key role in the superconductivity. Since the Y 5s orbitals have negligible DOS, the superlattice may be a three-band superconductor. Finally, the HSE06 hybrid functional was also used to optimize the structure and calculate the band structure of the superlattice SL1. Our results show that the electronic structure does not change a lot at the Fermi level as shown in Fig.~4 of the SI, and thus the pure DFT results are reliable here.

Besides the phonon spectra, Fig.~\ref{elph} also presents the electron-phonon coupling (EPC) of the strainless bulk YSi$_2$ and the YSi$_2$/Si superlattice SL1. In the bulk case, the lowest branch of the harmonic phonon spectra has small negative frequencies as mentioned before, which can be removed when the anharmonicity is included by using the SSCHA package. Apart from this, the anharmonicity also modifies remarkably the phonon dispersion and PHDOS of the higher branches. For example, the anharmonicity raises the peak of the Si PHDOS around $\omega=30$~meV. Correspondingly, the Eliashberg spectral function $\alpha^2F(\omega)$ also has remarkably larger value in the anharmonic case. Since $\alpha^2F(\omega)$ is derived by summing the EPC strength $\lambda_{n\mathbf{k}}$ across the entire Brillouin zone, the overall EPC also becomes stronger in the anharmonic case. As depicted in the Fig.~\ref{elph}(a), the EPC constant $\lambda$, which is roughly an integral of $\alpha^2F(\omega)$ over $\omega$, reaches approximately 0.744 under the harmonic approximation, and this value is comparable to that of MgB$_2$. After the anharmonicity is taken into account, $\lambda$ increases to about 1.05, which is so large that the Migdal-Eliashberg theory should be used to deal with its superconductivity. As for the YSi$_2$/Si superlattices, for example, SL1 shown in Fig.~\ref{elph}(b), the anharmonicity only increases $\lambda$ slightly from 0.6 to 0.7, which is smaller than the bulk value. The results for SL2 and SL3 are  given in the SI.

Now we are ready to present and discuss the key results for the superconductivity. The superconducting $T_\mathrm{c}$ of the bulk YSi$_2$ with and without strain were calculated by three methods: the McAD formula and the isotropic (anisotropic) ME theory with (without) anharmonicity. The three methods are implemented in Quantum Espresso (QE), SSCHA, and the Electron-Phonon Wannier (EPW), respectively. The YSi$_2$/Si superlattices were only studied by the McAD formula and isotropic ME theory with anharmonicity, because their EPW computation is too demanding.

\begin{table*}
\caption{\label{Tc-table-sl}The superconducting transition temperature $T_\mathrm{c}$ of the three YSi$_2$/Si superlattices calculated by using three methods.}
\begin{ruledtabular}
\begin{tabular}{ccccccc}
    \multirow{2}{*}{\textbf{Superlattice}} & \multirow{2}{*}{\textbf{q-grid}} & \textbf{Coarse} & \textbf{Fine} & \textbf{Harmonic} & \textbf{Anharmonic} & \textbf{Anharmonic} \\
     & & \textbf{k-mesh} & \textbf{k-mesh} & \textbf{McAD} & \textbf{McAD} & \textbf{Isotropic ME} \\ \hline
    SL1: 5L-YSi2/4L-Si & 5$\times$5$\times$2 & 10$\times$10$\times$2 & 20$\times$20$\times$2 & 4.2 & 8.1 & 9.1  \\ 
    SL2: 5L-YSi2/7L-Si & 5$\times$5$\times$2 & 10$\times$10$\times$2 & 20$\times$20$\times$2 & 3.4 & 6.3 & 7.1 \\
    SL3: 8L-YSi2/4L-Si & 5$\times$5$\times$1 & 10$\times$10$\times$1 & 20$\times$20$\times$1 & 6.0 & 14.3 & 15.4\\
\end{tabular}
\end{ruledtabular}
\end{table*}

The strainless bulk YSi$_2$ may have the highest theoretical $T_\mathrm{c}$, i.e., 20.6~K, in disilicides up to now. The McAD formula embedded in the QE package predicts $T_\mathrm{c}=8.1$~K, which is calculated with the q-mesh of 6$\times$6$\times$6 as shown in Table~\ref{Tc-table}. Because the EPC constant $\lambda$ is large enough ($\sim1$) to require a treatment based on the ME theory, we solved the anisotropic ME equation using the EPW package without the anharmonicity. The resulting $T_\mathrm{c}$ reaches 20.6~K as shown in Fig.~\ref{Tc}(a). This figure also shows that the strainless bulk YSi$_2$ has a single superconducting energy gap, althouth it is a three-band superconductor. This characteristic is quite different from that of MgB$_2$, which has two energy gaps. 

The anharmonicity also has remarkable effects on the superconductivity of YSi$_2$. Because the anharmonicity modifies the phonon spectra and electron-phonon coupling remarkably, we used the SSCHA package to take into account the anharmonicity and solve the isotropic ME equation for more accurate $T_\mathrm{c}$. When including the anharmonicity, the McAD formula now predicts a much higher $T_\mathrm{c}=14.2$~K, and the isotropic ME theory yields a higher $T_\mathrm{c}=15.3$~K as shown in Fig.~\ref{Tc}(b) and Table~\ref{Tc-table}. It is reasonable that $T_\mathrm{c}$ would be even higher when the solution of the anisotropic ME equation was implemented in the SSCHA code. We also tested the effects of $\mu^\ast$ on $T_\mathrm{c}$, and the results show that $T_\mathrm{c}$ varies from 12.7~K to 17.0~K as $\mu^\ast$ decreases from 0.16 to 0.07 as shown in Fig.~\ref{Tc}(b).

The in-plane compressive strain has complicated effects on $T_\mathrm{c}$ of the bulk YSi$_2$. We take one specific structure as an example: the in-plane lattice constant $a$ is compressed to $3.91$~Å and the out-of-plane lattice constant $c$ increases freely to $4.04$~Å so that $C_{66}$ is positive even without including the ZPE. The McAD formula in the QE predicts $T_\mathrm{c}=8.1$~K, which is nearly equal to that of the strainless case. When the anharmonicity is considered, the McAD formula and the isotropic ME equation yield $T_\mathrm{c}=13.6$~K and $T_\mathrm{c}=14.2$~K, respectively, which are smaller than those of the strainless case. This shows that the effects of the anharmonicity on $T_\mathrm{c}$ become weaker when the in-plane compressive strain makes the structure more stable mechanically. Then the solution to the anisotropic ME equation without the anharmonicity gives us $T_\mathrm{c}=14.1$~K. This suggests that the influence of the anisotropy on $T_\mathrm{c}$ also becomes weaker in the presence of the in-plane compressive strain. Finally, the feature of multi-gap superconductivity becomes discernible when an in-plane strain is applied to the bulk YSi$_2$ as shown in Fig.~5 of the SI. This suggests that the application of strain may be an effective way to engineer the character of superconducting energy gap.

When YSi$_2$ forms superlattices with Si, they might still keep relatively high $T_\mathrm{c}$ in silicides, although the interfaces between them decrease $T_\mathrm{c}$ further in addition to the strain effects discussed above . For example, when the McAD formula in the QE is used, $T_\mathrm{c}$ of the simplest structure, SL1, drops to around 4.3~K, which is still slightly higher than the boiling point of liquid helium. This is reasonable, becuase the Si honeycombs, which are inside the YSi$_2$ part but close to the interfaces, also become buckled somehow and less superconductive than the flat ones. When the anharmonicity is included, the McAD formula and the isotropic ME equation yield $T_\mathrm{c}=8.1$~K and $T_\mathrm{c}=9.1$~K, respectively. We also tested the effects of $\mu^\ast$ on $T_\mathrm{c}$, and the results show that $T_\mathrm{c}$ varies from 6.9~K to 10.6~K as $\mu^\ast$ decreases from 0.16 to 0.07 as shown in Fig.~\ref{Tc}(c). When comparing the three superlattices listed in Table~\ref{Tc-table-sl}, one can find that thicker YSi$_2$ tends to have higher $T_\mathrm{c}$, and it can almost reach the level of the bulk YSi$_2$ when the superlattice includes roughly 8 layers of YSi$_2$. Last but not least, the single-gap superconductivity of the bulk YSi$_2$ may become a multi-gap one due to the in-plane strain as well as the YSi$_2$/Si interfaces of the superlattices.~\cite{Bekaert2017} Theoretically, this could be tested by solving the anisotropic ME equation. This is a very interesting and challenging question that should be delt with seriously in the next steps as the computation is extremely demanding.

\section{Summary}

We proposed a Si-matched superconductor YSi$_2$ as an alternative to Al, which is usually the superconductor deposited on Si wafer in the mainstream superconducting quantum computers. On one hand, the anisotropic ME theory without the anharmonicity predicts that the strainless bulk AlB$_2$-type YSi$_2$ may have the highest $T_\mathrm{c}$ ($\sim20.6$~K) in the known disilicides, and this value is much higher than that of the bulk Al (1.14~K). When the remarkable anharmonicity is taken into account, even the simple McAD formula yields $T_\mathrm{c}=14.2$~K, much higher than the corresponding harmonic result $8.1$~K. The solution of the isotropic ME equation further raises $T_\mathrm{c}$ to 15.3~K. The anharmonicity can also remove the slightly negative frequencies in its phonon spectra, demonstrating its dynamical stability. These characteristics arise mainly from the `soft' Si honeycombs, which become plane in YSi$_2$ due to the `chemical flattening' effects of the Y atoms. Its thermodynamical and kinetic stability are supported by the convex hull and AIMD simulation results, respectively. Its mechanical stability is also verified by the calculation of the elastic stiffness tensor, especially, the most subtle term $C_{66}$, which is indeed positive if the ZPE correction is included in the energy-strain method and the k-mesh is dense enough. On the other hand, we studied the YSi$_2$(0001)/Si(111) superlattices to demonstrate the atomic-level compatibility of YSi$_2$ with the Si substrate and also introduce a new interpretation to the difference between theoretical and experimental lattice constants. All of the three superlattices demonstrate excellent kinetic stability, and strengthen the dynamical and mechanical stability of YSi$_2$. Meanwhile, they all keep $T_\mathrm{c}$ above 7~K, and can almost reach the level of the bulk YSi$_2$ when the layer number of YSi$_2$ is large enough (e.g., 8 layers). Moreover, the formation of the YSi$_2$/Si interfaces applies an in-plane compressive stress on YSi$_2$, and shortens (elongates) its lattice constant $a$ ($c$), which makes them close to the experimental results. Consequently, YSi$_2$ may demonstrate potential application value in quantum circuits, including Si-based JJFET and quantum computers.

\begin{acknowledgments}
This work was supported by the National Natural Science Foundation of China under Grant No. 12404258, and Hubei Provincial Key Research and Development Program under Grant No. 2025BAB034.
\end{acknowledgments}

\appendix

\section{Computation Methods}

\subsection{\label{structure-stability}Structure and Stability}

Utilizing density functional theory (DFT) as implemented in the Vienna Ab initio Simulation Package (VASP)~\cite{kresseVASP1996}, we carried out three types of computational tasks: structural optimization, calculation of elastic stiffness constants and \textit{ab initio} molecular dynamics (AIMD) simulations. Electron-ion interactions were modeled through projector augmented wave (PAW) potentials~\cite{PAW}, with valence configurations defined as 3s$^2$3p$^2$ for Si and 4s$^2$4p$^6$5s$^2$4d$^1$ for Y. The revised Perdew-Burke-Ernzerhof functional for solids (PBEsol) under generalized gradient approximation (GGA)~\cite{GGA-PBE-XC/PBE} was adopted to handle exchange-correlation interactions. Computational parameters included a 600 eV kinetic energy cutoff and a Monkhorst-Pack k-mesh of $12\times12\times8$ for Brillouin zone sampling during structural refinements and electronic convergence processes.~\cite{MP-Kmesh} Convergence thresholds were maintained at 10$^{-9}$~eV for energy differences, 10$^{-4}$~eV/Å for atomic forces, and 10$^{-3}$~GPa for stresses. 

The NVT ensemble with Nosé algorithm was used in the AIMD simulations.~\cite{NoseAlgorithm} For the bulk YSi$_2$, the simulations used a $5\times6\times4$ supercell containing 360 atoms. For the YSi$_2$/Si superlattices, the simulations used a $5\times6\times1$ supercell, since the out-of-plane dimension is aready large enough.

The thermodynamical stability of the bulk YSi$_2$ was tested by the method of convex hull. We considered both the electronic and phonon contributions to the energy. The electronic part of the energy was taken as the free energy of the relaxed structures, which includes the contribution of entropy.  In particular, we took into account the phonon part of the energy when we calculated all of the energies of the structures as shown in Fig.~\ref{stability}(b). The phonon part of the energy was taken as the ZPE, which is the sum of the energies of all phonon modes in the first dynamical matrix divided by two.

The elastic stiffness constants were calculated by using the energy-strain approach. A dense k-point mesh, e.g., on the level of $20\times20\times20$, must be used in order to get accurate results. When the ZPE needed to be taken into account, QE was used since all of the phonon spectral were calculated by this code. Otherwise, VASP was used in combination with the VASPKIT tool.~\cite{vaspkit}

\subsection{\label{band-phonon}Electronic Structure and Harmonic Phonon Spectra}

This investigation employed the Quantum ESPRESSO (QE) package~\cite{QE1,QE2,QE3} for the electronic structure, and the harmonic phonon spectra calculations. Specifically, we used the PBEsol functionals~\cite{GGA-PBE-XC/PBE} and the optimized norm-conserving Vanderbilt (ONCV) pseudopotentials~\cite{ONCV} to simulate the ion-electron interactions with valence configurations specified as 3s$^2$3p$^2$ for Si and 4s$^2$4p$^6$5s$^2$4d$^1$ for Y. The structural optimization continued until the atomic forces fell below 10$^{-15}$ Ry/Bohr, supported by plane-wave basis sets with the kinetic energy cutoff of 100~Ry for wavefunctions.

The lattice dynamics computations were performed within the framework of density functional perturbation theory (DFPT) as implemented in the PHonon module of QE. We used a q-mesh sampling up to 6$\times$6$\times$6 and carried out thorough convergence validation. Moreover, we employed the crystal acoustic sum rule (ASR) for the acoustic modes at the center of the Brillouin zone. All computational parameters underwent systematic verification through multiple convergence test cycles to ensure numerical reliability.

\subsection{Anharmonic Phonon spectra}

Anharmonic phonon spectra were calculated by using the stochastic self-consistent harmonic approximation (SSCHA).~\cite{SSCHA1,SSCHA2,SSCHA3,SSCHA4} Taking the bulk YSi$_2$ as an example, we carried out these computations using constant-volume relaxation protocols in 3$\times$3$\times$3 supercells in combination with $4\times4\times4$ k-grids. Free energy minimization involved iterative adjustments of atomic coordinates and force constant matrices, initialized using equilibrium atomic positions of DFT and DFPT-derived dynamical matrices calculated on a 3$\times$3$\times$3 q-grid with 100~Ry plane-wave cutoff. Simulations maintained a 0~K thermal environment until the termination criteria were fulfilled. Then the SSCHA phonon dispersion relations and polarization vectors were derived from the Hessian of the minimized free energy. 

\subsection{Electron-Phonon Coupling and Superconductivity}

The fundamental electron-phonon coupling (EPC) properties and superconductivity were calculated by following the standard procedure of QE. First, we used the PHonon module (specifically ph.x) to calculate the phonon linewidth $\gamma_{\mathbf{q}\nu}$ and EPC strength $\lambda_{\mathbf{q}\nu}$ for mode $\nu$ at wavevector $\mathbf{q}$ using the double-delta function approximation. Second we transformed the electron-phonon coefficients into the real space (using q2r.x). Then we calculated the EPC strength and the spectral function $\alpha^2F$ on a much denser k-mesh (using matdyn.x). Finally, we calculated the total EPC constant $\lambda$ and its corresponding superconducting $T_\mathrm{c}$ by the McAD formula (using lambda.x).

On the basis of the QE results, the EPC and superconductivity were studied more accurately by using EPW.~\cite{Wannier-Interpolation1/EPW} On one hand, the phonon linewidth, which corresponds to the imaginary part of the phonon self-energy, could be calculated without the double-delta function approximation when using EPW. The EPC strength $\lambda(\mathbf{k},\mathbf{k}',n-n')$ and Eliashberg spectral function $\alpha^2F(\mathbf{k},\mathbf{k}',\omega)$ now become anisotropic. On the other hand, superconducting parameters were determined by the numerical solutions of the anisotropic Migdal-Eliashberg equations~\cite{ME1,ME2,ME3}. The solutions could give us the superconducting energy gap $\Delta(\mathbf{k},\omega)$ as a function of temperature, and enable us to see if the superconductor is a single-gap or multi-gap one. However, the anharmonicity has not been taken into account by the solution of the anisotropic ME equation yet.

The anharmonicity were taken into account by the SSCHA code. First, we used QE to calculate the harmonic dynamical matrices on a q-mesh of suitable size. Specifically, for the bulk YSi$_2$, we chose a 3$\times$3$\times$3 q-mesh, which later instructs SSCHA to create 3$\times$3$\times$3 supercells. Second, the SSCHA code took the harmonic dynamical matrices as inputs and yielded dynamical matrices with anharmonicity as outputs. Third, the anharmonic dynamical matrices are transformed to the force constants in the real space (using q2r.x). Fourth, the SSCHA code used a modified version of QE to calculate the EPC strength, the spectral function $\alpha^2F$, and $T_\mathrm{c}$ using McAD formula on the basis of the anharmonic force constants. Finally, SSCHA can also solve the isotropic ME equation to give us more accurate $T_\mathrm{c}$. However, SSCHA cannot solve the anisotropc ME equation yet.

\bibliography{ref.bib}

\end{document}